\documentclass[12pt,thmsa]{article}
\usepackage{amsfonts}
\usepackage{graphicx}
\usepackage{epsfig}
\usepackage{graphics}
\usepackage{setspace}
\makeatletter
\newcommand{\row}[1]%
{\mathord{\buildrel{\lower3pt%
\hbox{$\scriptscriptstyle\rightarrow$}}\over #1}}
\newcommand{\col}[1]{{#1^{\raisebox{2pt}[\height]%
{$\scriptstyle\downarrow$}}}}
\newcommand{\dyadic}[1]{\mathord{\dyadic@rrow{#1}}}
\newcommand{\dyadic@rrow}[1]{
\begin{picture}(12,12)(-1,0)
\put(-1,9){\makebox(0,0)[t]{$\scriptscriptstyle\downarrow$}}
\put(-1,9){\makebox(0,0)[l]{$\scriptscriptstyle\longrightarrow$}}
\put(5,0){\makebox(0,0)[b]{$#1$}}
\end{picture}
}

\newcommand{\bra}[1]{\bigl\langle #1 \bigr|}
\newcommand{\ket}[1]{\bigl| #1 \bigr\rangle}

\begin{document}

\begin{center}
{\large  Estimation of teleported and gained parameters in a  non-inertial frame}\\
 {N. Metwally\\}
$^1$ Department of Mathematics, College of Science, Bahrain
University, Bahrain \\
$^2$Department of Mathematics, Faculty of Science,
Aswan University, Aswan, Egypt \\
\end{center}
\date{\today }

\begin{abstract}
We estimate the teleported and the gained parameters by means of
Fisher information in a non-inertial frame. The sender and the
receiver share an accelerated maximum or partial entangled state.
The estimation degree of  these parameters depends on Unruh
acceleration, the used single mode approximation(within/ beyond),
the structure of the initial teleported state, and the
entanglement of the initial communication channel. The
maximization and minimization  estimation degree of the teleported
parameters depend on the initial encoded information. The
estimation degree of the parameters can be maximized if the
partners teleport classical information.

\end{abstract}
{\bf keywords}: Estimation, Fisher information, Unruh acceleration
%\maketitle

%\tableofcontents
\section{Introduction}
In the context of the classical estimation theory, Fisher
information represents a key value that used to estimate any
parameter \cite{Holevo, Wang}. Similarly, quantum Fisher
information can be used as a measure of estimation   in quantum
information theory \cite{bran}. As an example, the  optimal
parameter estimation of Pauli channels is discussed by Ruppert et.
al., \cite{Ruppert}.   Q. Zheng et. al, \cite{Zheng} have used
Fisher information to estimate the channel parameter of a two-
qubit state, where each qubit interacts independently with its own
environment. The possibility of estimating  multi-quantum
parameters  is discussed  by P. Yue  et. al, \cite{Ping}.

 Due to its important in estimation theory, there are many studies
 that have quantified  the Fisher information for different models. For
 example, quantum Fisher information of the GHZ state in
 decoherence channel is quantified by J. Ma et. al {\cite{Jian}.
 The relation between the fidelity susceptibility and
quantum Fisher information is investigated  by J. Liu et. al
\cite{Jing}. The dynamic of the quantum Fisher information in the
 Ising model is discussed in\cite{Heng}. A. Altintas \cite{Ali}
 studied the dynamics of  quantum Fisher information of a steady
 state in a noisy environment. Recently, J. He et.al,
 \cite{Juan} studied the possibility of
 enhancing the quantum Fisher information by using the
 uncollapsing measurements. Xiao et. al \cite{Xing} proposed a
 scheme to enhance the teleported quantum Fisher information  by
 utilizing partial measurements.

However, there are limited studies have been done in the context
of non-inertial frame. For example, Yao et. al, \cite{Yao} have
investigated the performance of the quantum  Fisher information
under the Unruh-Hawking effect. The dynamics of Fisher information
and skew information for Unruh  effect within and witout external
noise is discussed by Banerjee et. al \cite{Ban}. Recently,
Metwally \cite{Metwally2016} discussed the effect of the Unruh
acceleration on the Fisher information for different classes of
maximum and partial entangled state.

Therefore, we are motivated to  discuss the possibility of
estimating the teleported parameters by means of teleporting a
quantum state between two users. In this proposal, it is assumed
that one user (Alice) stays in inertial frame, while the second
user (Bob) is accelerated with a uniform acceleration.

 The manuscript is
organized as follows: in Sec.$2$, we describe   different types of
the initial states that  Alice and Bob share, the relation between
the Minkowski and Rindler spaces is reviewed  , and the dynamics
of the final state when only Bob's qubit is accelerated uniformly
is obtained analytically.  This state is used as a quantum
communication  channel to teleport unknown state from Alice to Bob
as described in Sec.$3$.  In Sec.4, we estimate the teleported
parameters by means of Fisher information. Finally, we summarize
our results in Sec.(5).

\section{The suggested Proposal }
 We  assume that, Alice and Bob as  partners share
a state prepared initially in a two qubit system of Bell, or
$X$-state. It is considered that, only one qubit say, Alice's
qubit is at  a rest, while Bob's qubit is accelerated with a
uniform Unruch acceleration.  Alice's task is teleporting unknown
 state to Bob by using Bennett's protocol \cite{Bennt,Deutsch}. Bob will use
 this teleported state to estimate the initial parameters of the
 teleported state and the gained parameters during the
 teleportation process.  Bob will then quantify the Fisher
 information \cite{bran} corresponding to these parameters.
 In this context, we aim to investigate the
effect of the Unruch acceleration, and  the initial state settings
of the  communication channel between the partners, on the
precision of estimation.

Let  the partners, Alice and Bob share  a self-transposed class of
two qubits \cite{Englert1} as,

\begin{equation}\label{ini}
\rho_{12}=\frac{1}{4}\Bigl(I_{4\times4}+
\sum_{i,j}\row{\sigma_i^{(1)}}\cdot\dyadic{C}\cdot\row{\sigma_j}^{(2)}\Bigr),
\end{equation}
where
$\row{\sigma^{(i)}}=(\sigma_x^{(i)},\sigma_y^{(i)},\sigma_z^{(i)})$,
 are  the Pauli
matrices of  Alice's and Bob's qubits, respectively. The dyadic
$\dyadic{C}$ is a $3\times 3$ matrix, where its elements are
defined as $c_{ij}=tr\{\rho_{12}\sigma_i^{(1)}\sigma_j^{(2)}\}$
\cite{Englert1}. From the state (\ref{ini}),  different important
states can be considered. However if we set $c_{11}\neq c_{22}\neq
c_{33}\neq 0$ and $c_{ij}=0,i\neq j$ one obtains what is called
$X$- state. Moreover, if $c_{11}=c_{22}=c_{33}=-1$, and $c_{ij}=0,
i\neq j$  one gets a maximum Bell state, singlet state
$\rho_{\psi^-}$. Also, what is called Werner state can be obtained
if $c_{11}=c_{22}=c_{33}=-F$ and $c_{ij}=0, i\neq j$. The dynamics
of all these states in the non-inertial frame is discussed  by
Metwally \cite{Nasser2013}.

\subsection{ Unruh effect}

It has been shown that, for the perspective inertial observers the
Minkowski coordinates are the most suitable choice to describe the
Dirac qubit. On the other hand, from the scenery of the
non-inertial observers, the Rindler coordinates are the most
adequate coordinates to describe the Dirac qubits. Now,  we assume
that, Alice's qubit moves in the inertial frame,  and described by
Minkowski coordinates $(t, z)$, while Bob's qubit is uniformly
accelerated  with a constant acceleration $a$. Therefore, Bob's
qubit can be described by using Rindler coordinates $(\tau,\eta)$,
such that, $\tau=tanh^{-1}(t/x)$, $\eta=\sqrt{x^2-t^2}$,
$0<\eta<\infty$, $-\infty<\tau<\infty$. These transformations
define two regions in the space-time; the first region, $I$ for
$|t|<z$ and the second region, $II$ for $z<-|t|$. The accelerated
qubit moves on a parabola in the first region $I$ defined by
$\eta=1/a$, $a$ is the uniform acceleration, while the
Anti-accelerated qubit moves on the parabola $\eta=-1/a$ in the
second region $II$ \cite{Als,Edu}.
 To describe a
Minkowski sate in terms of Rindler's space, one has to use the
Bogoliubov transformation,
\begin{equation}
\nu_k=\cos r\mathcal{C}_k^I-e^{-i\phi}\sin r
\mathcal{D}_k^{II},\quad \mu_k^\dagger=e^{i \phi}\sin
r\mathcal{C}_k^I+\cos r \mathcal{D}^{II}_k,
\end{equation}
where $\nu_k$ and $\mu^\dagger_k$ represent the annihilation and
criterion operators in Minkowski space such that \cite{Als},
\begin{equation}
\nu_k\ket{0_k^+}_M=0,\quad \mu_k\ket{0_k^-}_M=0,\quad
\nu^\dagger_k\ket{0_k^+}_M=\ket{1_k^+},\quad
\mu^\dagger_k\ket{0_k^-}_M=\ket{1_k^-}_M,
\end{equation}
and, $\mathcal{C}^I_k$ and  $\mathcal{D}^{II^\dagger}_{-k}$
represent the annihilation and creation operators in the regions
$I$ and $II$, respectively. The parameter $\phi$ is an unimportant
phase that can be absorbed into the definition of the operators.

\begin{eqnarray}\label{trans}
\ket{0_k}&=&\cos r_b\ket{0^+_k}_I\ket{0^{-}_{k}}_{II}+ \sin
r_b\ket{1^{+}_k}_I\ket{1^{-}_{k}}_{II}, \nonumber\\
\ket{1_k}&=&q_R\ket{1^{+}_k}_I\ket{0^{-}_k}_{II}+q_L\ket{0^{+}_k}_I\ket{1^{-}_k}_{II},
\end{eqnarray}
where $q_R$ and $q_L$ are complex numbers with
$|q_R|^2+|q_L|^2=1$,  and the dimensionless  parameter $r$ is
given from
 $tan r=e^{-\pi\omega \frac{c}{a}}$, $a\in[0,\infty)$, $r\in[0,\pi/4]$,  $\omega$ is the frequency of the
travelling qubits, and  $c$ is the speed of light.  The
transformation (\ref{trans}) mixes a particle (fermions) in region
$I$ and an anti particle (anti-fermions) in region $II$. In the
computational basis $\ket{0_k}$and $\ket{1_k}$ can be written as
\cite{Edu,un1}.

\subsection{Accelerating the communication quantum state}

Let us consider that, the partners Alice and Bob share a state
defined by (\ref{ini}).  As  described  above,  Alice's qubit
remains stationary, while Bob's qubit is accelerated with  a
uniform acceleration. Bob is causally disconnected from the second
region, therefore all  the accessible information is encoded in
Alice's
 and Bob's qubit in the first region. Consequently, if we
trace out all the modes in the second region, the final state
between Alice and Bob is given by,
\begin{eqnarray}\label{acc}
\rho_{ab}^{acc}&=&\mathcal{B}_1\ket{00}\bra{00}+\mathcal{B}_2\ket{00}\bra{11}+\mathcal{B}_3\ket{11}\bra{00}
+\mathcal{B}_4\ket{01}\bra{01}
\nonumber\\
&&+\mathcal{B}_5\ket{10}\bra{10}+\mathcal{B}_6\ket{10}\bra{01}+\mathcal{B}_7\ket{01}\bra{10}
+\mathcal{B}_8\ket{11}\bra{11},
\end{eqnarray}
where,
\begin{eqnarray}
\mathcal{B}_1&=&\mathcal{A}_1\cos^2r+\mathcal{A}_2|q_L|^2,\quad
\mathcal{B}_2=\mathcal{A}_4 q^*_R\cos r+\mathcal{A}_3 q_L\sin r,
\nonumber\\
\mathcal{B}_3&=&\mathcal{A}_3q^*_L\sin r+\mathcal{A}_4 q_R\cos
r,\quad
 \mathcal{B}_4=\mathcal{A}_1\sin^2r+\mathcal{A}_2|q_R|^2,
 \nonumber\\
 \mathcal{B}_5&=& \mathcal{A}_2\cos^2r+\mathcal{A}_1|q_L|^2,\quad
 \mathcal{B}_6=\mathcal{A}_3q^*_R\cos r+\mathcal{A}_4 q_L\sin r,\quad
 \nonumber\\
 \mathcal{B}_7&=&\mathcal{A}_3q^*_L\sin r+\mathcal{A}_3 q_R\cos r,\quad
 \mathcal{B}_8= \mathcal{A}_2\sin^2r+\mathcal{A}_1|q_R|^2,
\end{eqnarray}
and $|q_r|^2+|q_L|^2=1,~ \mathcal{A}_1=\frac{1+c_{33}}{4},~
\mathcal{A}_2=\frac{1-c_{33}}{4},~
\mathcal{A}_3=\frac{c_{11}+c_{22}}{4}$ and
$\mathcal{A}_4=\frac{c_{11}-c_{22}}{4}$.

In the next section, Alice and Bob will use the state (3) as a
communication channel to teleport an unknown state from Alice to
Bob by using Bennett protocol \cite{Bennt}.

\section{Quantum Teleportation} Now, the  partners
share the accelerated state(\ref{acc}) and Alice is asked to send
the unknown state, $\rho_u$, to Bob where,
\begin{equation}
\rho_u=|\alpha|^2\ket{0}\bra{0}+\alpha\beta^*\ket{0}\bra{1}+\beta\alpha^*\ket{1}\bra{0}+|\beta|^2\ket{1}\bra{1},\quad
|\alpha|^2+|\beta|^2=1.
\end{equation}
 Alice and Bob perform the teleporation protocol\cite{Bennt} by using the
following steps.
\begin{enumerate}
\item Alice performs a CNOT operation between her own qubit and
the given one followed by the Hadamard gate on the given qubit.

\item Alice performs  measurements on the two qubits on her hand
and send her results to Bob via classical communication channel.
\item According to the received results, Bob performs the required
operations to get the teleported state.

\end{enumerate}
 Alice and Bob perform the
steps(1-3)  to teleport the unknown state $\rho_u$. Finally,  if
Alice measures $00$, then Bob will obtain the state,
\begin{equation}\label{telp}
\rho_{Bob}=\varrho_{00}\ket{0}\bra{0}+\varrho_{01}\ket{0}\bra{1}+\varrho_{10}\ket{1}\bra{0}+\varrho_{11}\ket{1}\bra{1},
\end{equation}
where,
\begin{eqnarray}\label{coff}
\varrho_{00}&=&\frac{1}{2}(|\alpha|^2\mathcal{B}_1+|\beta|^2\mathcal{B}_5),\quad
\varrho_{01}=\frac{1}{2}(\alpha\beta^*\mathcal{B}_2+\beta\alpha^*\mathcal{B}_6),\quad
\nonumber\\
\varrho_{10}&=&\frac{1}{2}(\alpha\beta^*\mathcal{B}_7+\beta\alpha^*\mathcal{B}_3),\quad
\varrho_{11}=\frac{1}{2}(|\alpha|^2\mathcal{B}_4+|\beta|^2\mathcal{B}_8).
\end{eqnarray}
Let us assume that, the coefficients $\alpha=\cos(\theta/2)$ and
$\beta=\sin(\theta/2)e^{i\phi}$, where the parameters
$\theta\in[0,\pi]$ and $\phi\in[0,2\pi]$ are the weight and the
phase angles, respectively. Then (\ref{coff}) can be written
explicitly as,

\begin{eqnarray}
\varrho_{00}&=&\frac{1}{8}\left[\cos^2r(1+c_{33}\cos\theta)+|q_L|^2(1-c_{33}\cos\theta)\right],
\nonumber\\
\varrho_{01}&=&\frac{\sin\theta}{8}\Bigl[c_{11}\cos\phi(q^*_R\cos
r+q_L\sin r)+ic_{22}\sin\phi(q^*_R\cos r-q_L\sin r)\Big],
 \nonumber\\
\varrho_{01}&=&\frac{\sin\theta}{8}\Bigl[c_{11}\cos\phi(q^*_L\sin
r+q_R\cos r)+ic_{22}\sin\phi(q^*_L\sin r-q_R\cos r)\Big],
\nonumber\\
\varrho_{11}&=&\frac{1}{8}\left[\sin^2r(1+c_{33}\cos\theta)+|q_R|^2(1-c_{33}\cos\theta)\right].
\end{eqnarray}
Now, we have all  details to estimate the weight $(\theta)$, the
phase $(\phi)$ and the Unruh  $(r)$ parameters by calculating the
Fisher information corresponding to each parameter as we see in
the next sections.

\section{Fisher Information}
It is clear that, the final teleported state depends on the
initial parameters, the weight and the phase parameters as well as
the Unruh parameter which is gained during the teleportation
process. The main task of the following subsections is estimating
these parameters by calculating the Fisher information
corresponding to these parameters.

It is well known that, any  single mixed qubit can be described by
its Bloch vector as,
\begin{equation}
\rho=\frac{1}{2}(1+\row{s}\cdot\col{\sigma}),
\end{equation}
where $\row{s}=(s_x,s_y,s_z)$ and
$\col\sigma=(\sigma_x,\sigma_y,\sigma_z)^T$. Fisher information
for a mixed state with respect to a parameter $\kappa$, which will
be estimated, can be described by means of the Bloch vector as
\cite{Sun},
\begin{eqnarray}
\mathcal{F}_\kappa=\Big|\frac{\partial{\row{s}}}{\partial\kappa}\Big|^2+\frac{1}{1-|\row{s}|^2}
\Bigl(\row{s}\cdot\frac{\partial{\row{s}}}{\partial\kappa}\Bigr)^2,
\end{eqnarray}
while for pure state, namely $|\row{s}|=1$, the Fisher information
$\mathcal{F}_\kappa=\Big|\frac{\partial{\row{s}}}{\partial\kappa}\Big|^2$.

Now, to quantify the amount of the teleported Fisher information
which  contained in the state (\ref{telp}), we  describe it by
means of its Bloch vector as,
\begin{equation}\label{tBob}
\rho_{Bob}=\frac{1}{2}(1+s_x\sigma_x+s_y\sigma_y+s_z\sigma_z),
\end{equation}
where $s_i=Tr\{\rho_{Bob}\sigma_i\}, i=x,y$ and $z$,
\begin{eqnarray}
s_x&=&\frac{\sin\theta}{8}\Bigl[c_{11}\cos\phi\{(q^*_R+q_R)\cos
r+(q^*_L+q_L)\sin r\} \nonumber\\
&&+ic_{22}\sin\phi\{(q^*_R-q_R)\cos r+(q^*_L-q_L)\sin r\}\Big]
\nonumber\\
s_y&=&\frac{\sin\theta}{8}\Bigl[c_{22}\sin\phi\{(q^*_R+q_R)\cos
r-(q^*_L+q_L)\sin r\} \nonumber\\
&&+ic_{11}\cos\phi\{(q_R-q^*_R)\cos r+(q_L-q^*_L)\sin r\}\Big]
\nonumber\\
s_z&=&\frac{1}{8}\Bigl[\cos2r(1+c_{33}\cos\theta)+(|q_L|^2-|q_R|^2)(1-c_{33}\cos\theta)\Bigr].
\end{eqnarray}

\subsection{Estimation of the weight parameter, $\Large\theta$ }

In this investigation, we assume that the partners initially share
maximum entangled state of Bell type, $\rho_{\phi^+}$ or
$\rho_{\psi^-}$, partial entangled state of $X$-state. In this
context, we shall estimate the teleported weight and phase
parameters by calculating the Fisher information with respect to
these two parameters.

\begin{figure}[t!]
\centering
                   \includegraphics[width=15pc,height=15pc]{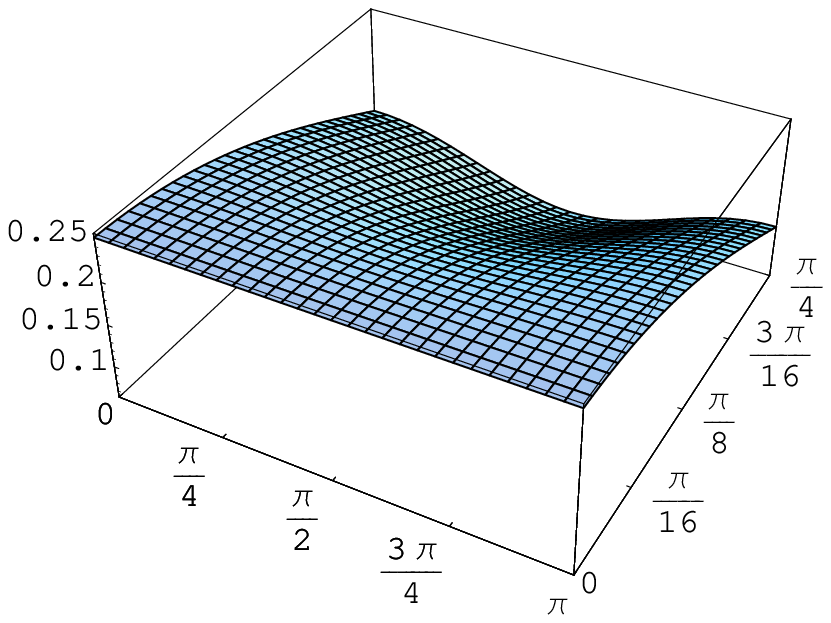}
           \put(-140,20){\Large$\theta$}
             \put(-10,60){\Large$r$}
      \put(-198,85){$\mathcal{F}_\theta$}
     \put(-170,160){$(a)$}~~\quad\quad
               \includegraphics[width=15pc,height=15pc]{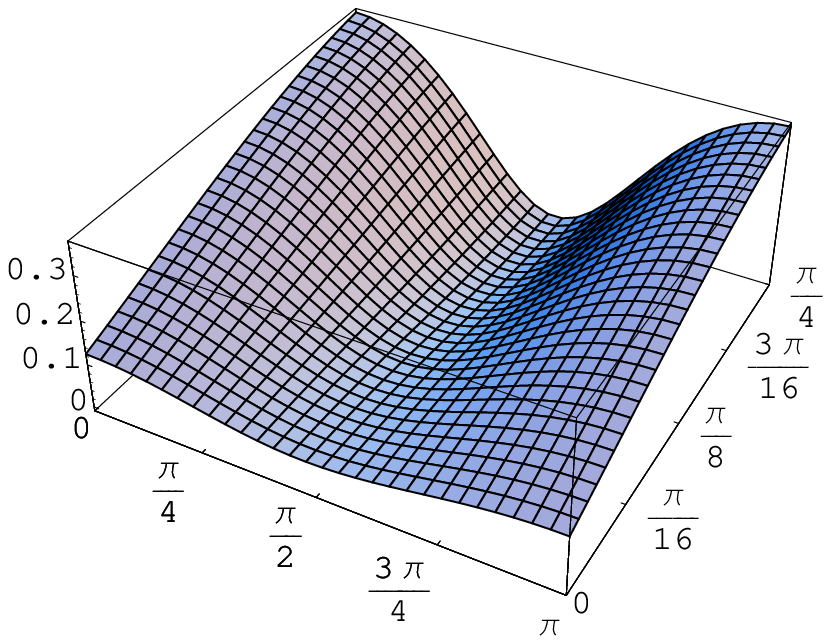}
          \put(-140,20){\Large$\theta$}
             \put(-10,60){\Large$r$}
     \put(-198,85){\Large$\mathcal{F}_\theta$}
      \put(-160,160){$(b)$}
             \caption{ The  dynamics of Fisher information $\mathcal{F}_\theta(\theta,r)$ at a fixed
             $\phi=\pi/4$ against the Unruh
            parameter, $r$: (a) WSMA, namely, $q_R=1$ and $q_L=0$, and (b)BSMA with $q_R=q_L=1/\sqrt{2}$.
             }
\end{figure}

\begin{figure}[t!]
\centering
     \includegraphics[width=15pc,height=15pc]{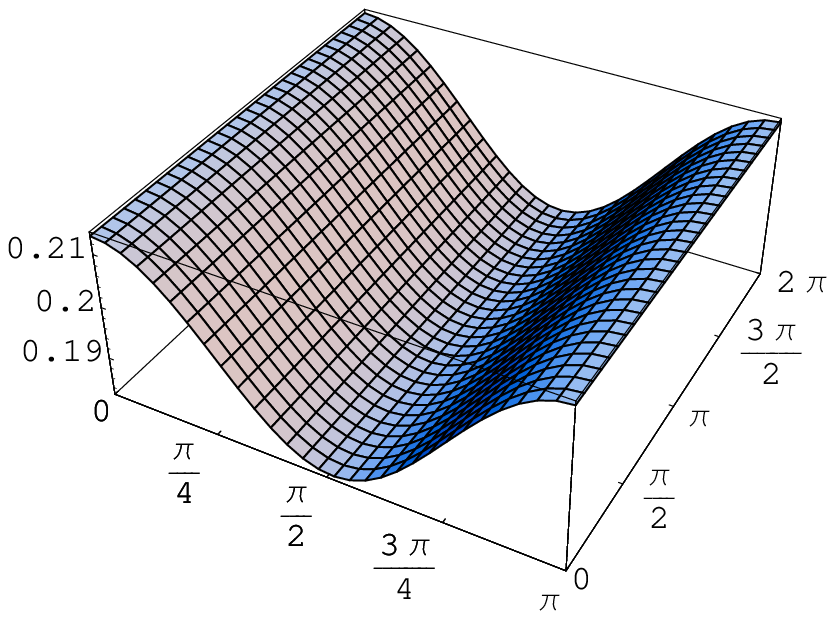}
      \put(-180,160){$(a)$}
    \put(-140,20){\Large$\theta$}
                \put(-15,55){\Large$\phi$}
               \put(-198,85){\Large$\mathcal{F}_\theta$}~~\quad\quad
   \includegraphics[width=14pc,height=14pc]{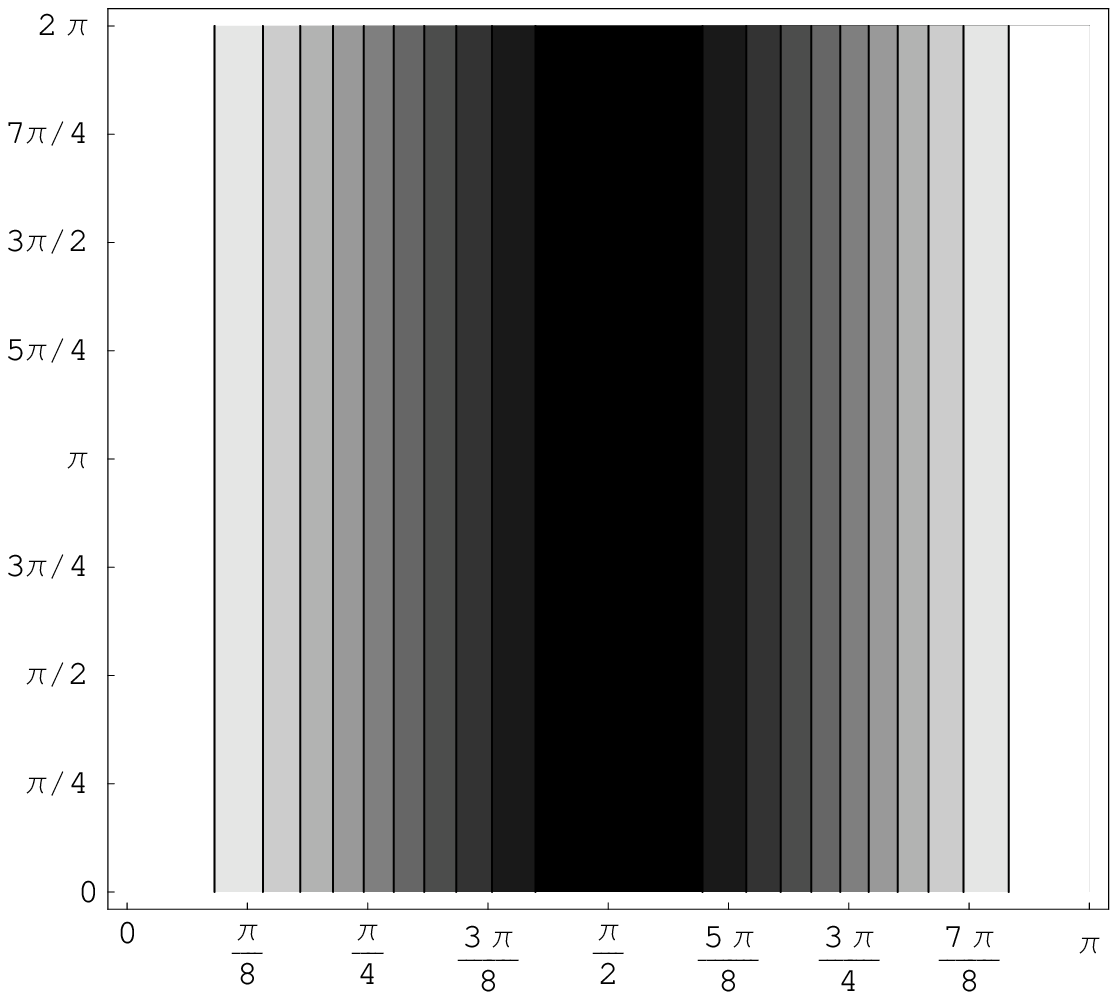}
                 \put(-180,85){\Large$\phi$}
                  \put(-80,-10){\Large$\theta$}
     \put(-175,160){$(b)$}\\
       \includegraphics[width=15pc,height=15pc]{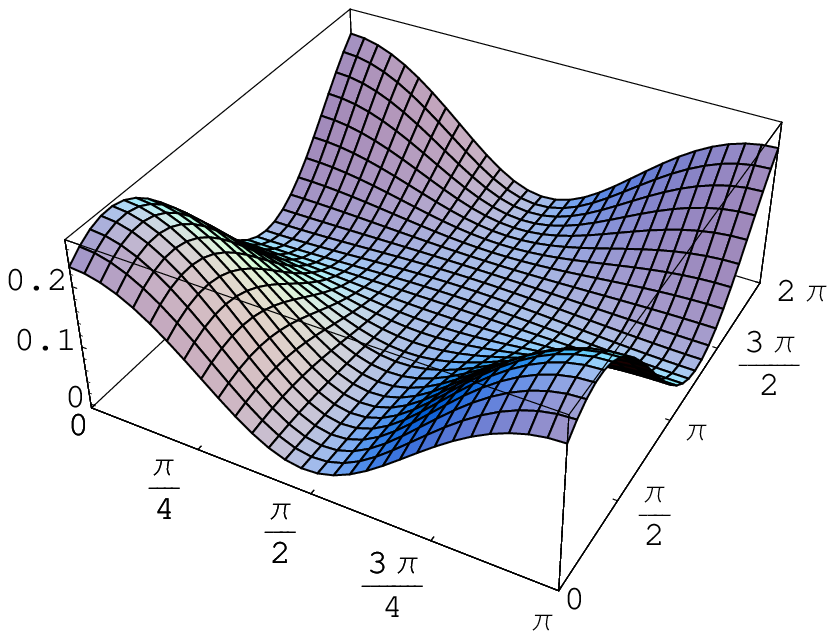}
       \put(-170,160){$(c)$}
       \put(-140,20){\Large$\theta$}
      \put(-15,55){\Large$\phi$}
      \put(-198,85){\Large$\mathcal{F}_\theta$}~~\quad\quad
       \includegraphics[width=14pc,height=14pc]{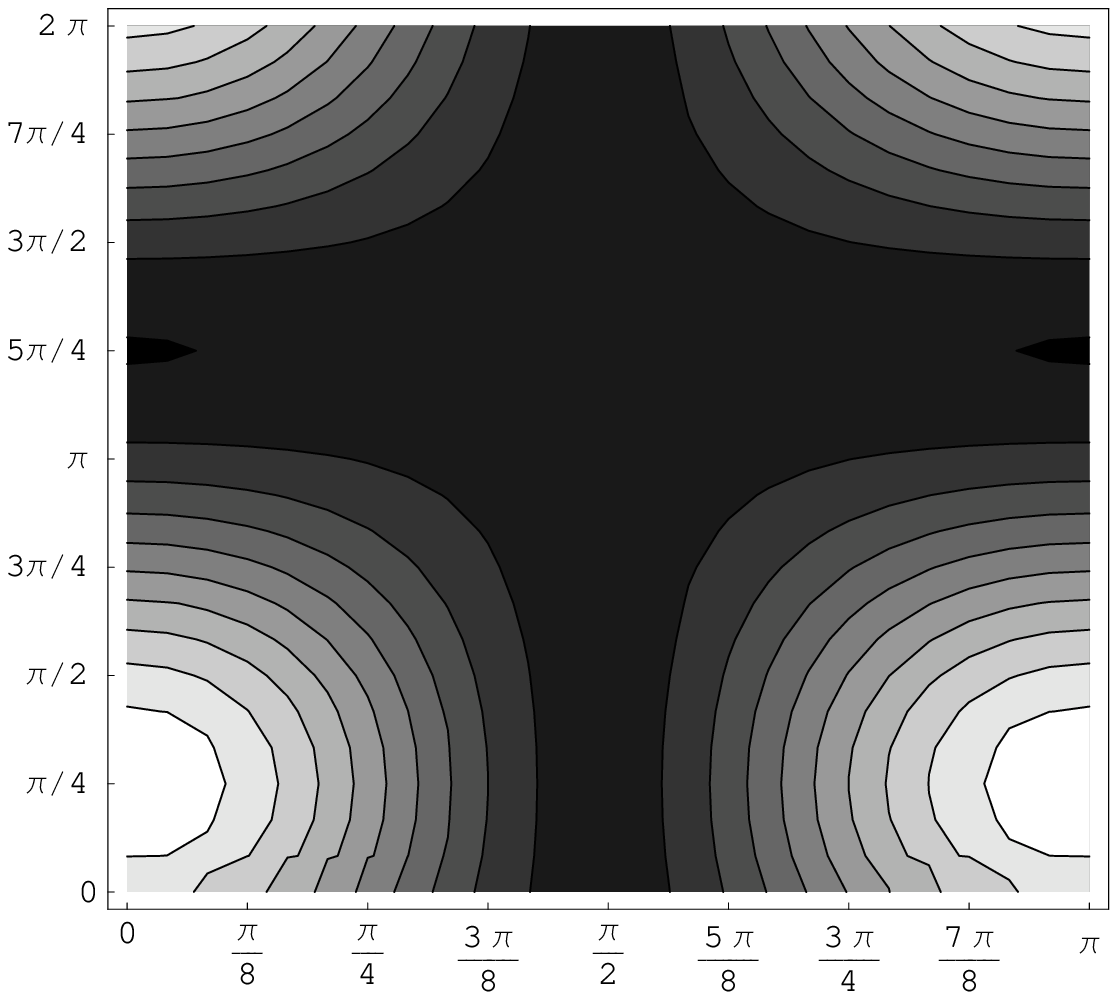}
                        \put(-80,-10){\Large$\theta$}
             \put(-180,85){\Large$\phi$}
      \put(-180,160){$(d)$}
            \caption{ The same as Fig.(1) but it represents the
            Fisher information, $\mathcal{F}_{\theta}(\theta,\phi)$  at a fixed $r=\pi/8$ is plotted
            against the phase  parameter $\phi$.}
\end{figure}

Fig.(1) displays the behavior of the Fisher information
$\mathcal{F}_\theta(\theta,r)$ at fixed $\phi=\pi/4$ with respect
to the weight parameter, $\theta$ of the teleported state
(\ref{tBob}) and the Unruh acceleration within and beyond the
single approximation.  In Fig.(1a), we  consider $q_R=1$ while
$q_L=0$, namely within the single mode approximation, (WSMA). The
general behavior shows that, Fisher information decreases as $r$
increases. The effect of the weight parameter $\theta$  appears
clearly for large values of $r$, where
$\mathcal{F}_\theta(\theta,r)$ decreases as $r$ increases to reach
its minimum values at $\theta=\pi/2.$ However, for further values
of $\theta$, the Fisher information increases gradually to reach
its maximum value at $\theta=\pi$.

This behavior   changes  dramatically when we consider the Unruh
effect beyond the single mode approximation,(BSMA), namely the
single qubit has  right and left components, where we set
$q_R=q_L=1/\sqrt{2}$. As it is  described in Fig.(1b), Fisher
information increases as $r$ increases.  Moreover, as $\theta$
increases, Fisher information decreases to reach its minimum
values at $\pi/2$. Then as  the weight parameter  increases,
$\mathcal{F}_\theta(\theta,r)$ increases gradually to reach its
maximum values. These maximum values depend on the Unruh
acceleration, where the maximization of
$\mathcal{F}_\theta(\theta,r)$ is depicted at large values of $r$.

The effect of the phase  parameter, $\phi$ on the dynamics of
Fisher information, $\mathcal{F}_\theta(\theta,\phi)$ at fixed
$r=\pi/8$ is depicted in Fig(2). The behavior of the Fisher
information within the single mode approximation is displayed in
Fig.(2a). It is clear that, the phase  parameter has a negligible
 effect on the behavior of $\mathcal{F}_\theta(\theta,\phi)$. On the
other hand, the phenomena of the sudden decay of the Fisher
information is displayed as soon as $\theta$ increases.
 However, at $\theta\simeq \pi/2$, $\mathcal{F}_\theta(\theta,\phi)$ vanishes completely.
For further values, Fisher information re-birth again to reach its
maximum value at $\theta=\pi$.   Fig.(2b) shows the behavior of
$\mathcal{F}_\theta(\theta,\phi)$, but in a contour description,
where it displays clearly, the values of $\theta$ at which Fisher
information decreases/increases and the values of $\theta$ that
maximize or minimize $\mathcal{F}_\theta(\theta,\phi)$.

Fig.(2c) displays the behavior of Fisher information
$\mathcal{F}_\theta(\theta,\phi)$ against the phase parameter,
$\phi$ BSMA, where we fixed the value of Unruh acceleration,
$r=\pi/8$. In this case, the phase parameter has a completely
different effect. It is evident that, as $\phi$ increases,
$\mathcal{F}_\theta(\theta,\phi)$ decreases gradually to reach its
minimum  values at $\phi=\pi$. For further values of $\phi$, the
Fisher information  completely vanishes to re-birth again for
$\phi\in[3\pi/4,2\pi]$. These results are displayed in Fig.(2d),
where the values of $\theta$ and $\phi$ which maximize and
minimize $\mathcal{F}_\theta$ are seen clearly.

 Figs.$(3a\&3b)$ are devoted to investigate the effect of the
phase  parameter, $\phi$  on the Fisher information,
$\mathcal{F}_\theta(\theta,\phi)$  BSMA at fixed $r=\pi/8$, where
the partners initial share the singled state, $\rho_{\psi^-}$. It
is evident that, $\mathcal{F}_\theta(\theta,\phi)$ increases as
$\phi$ increases in the interval $[0,\pi]$ and reaches its maximum
values at $\phi=3\pi/4$, then it decreases  again  at  further
values of $\phi$ to vanish completely at $\phi\in[3\pi/2,2\pi]$.
On the other hand, as $\theta$ increases,
$\mathcal{F}_\theta(\theta,\phi)$, decreases gradually to vanish
completely at $\theta\in[\pi/4,3\pi/4]$. For further values of
$\theta$, $\mathcal{F}_\theta(\theta,\phi)$ re-births again to
reach its maximum bounds at $\theta=\pi$.

\begin{figure}[t!]
\centering
         \includegraphics[width=15pc,height=15pc]{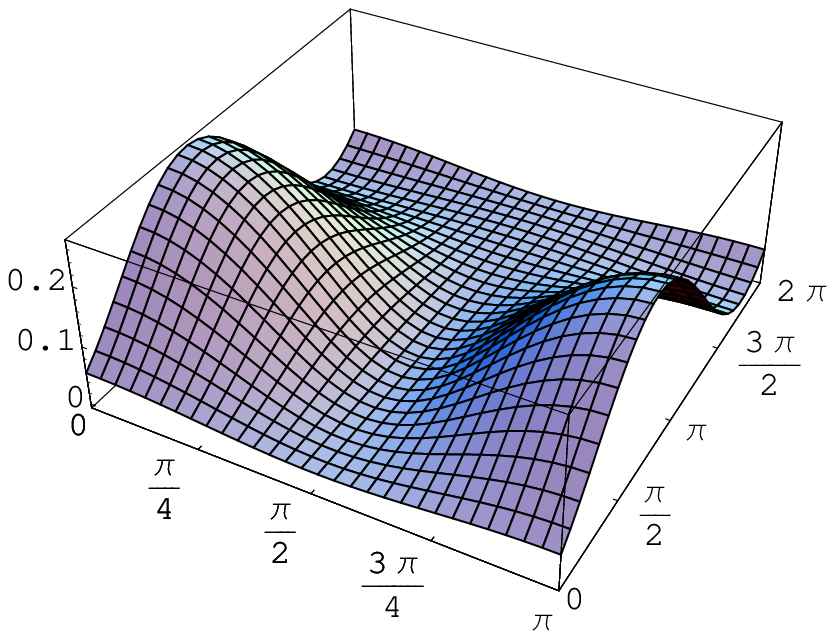}
          \put(-170,160){$(a)$}
         \put(-145,20){\Large$\theta$}
                \put(-15,55){\Large$\phi$}
               \put(-198,85){\Large$\mathcal{F}_\theta$}~\quad\quad\quad
           \includegraphics[width=14pc,height=14pc]{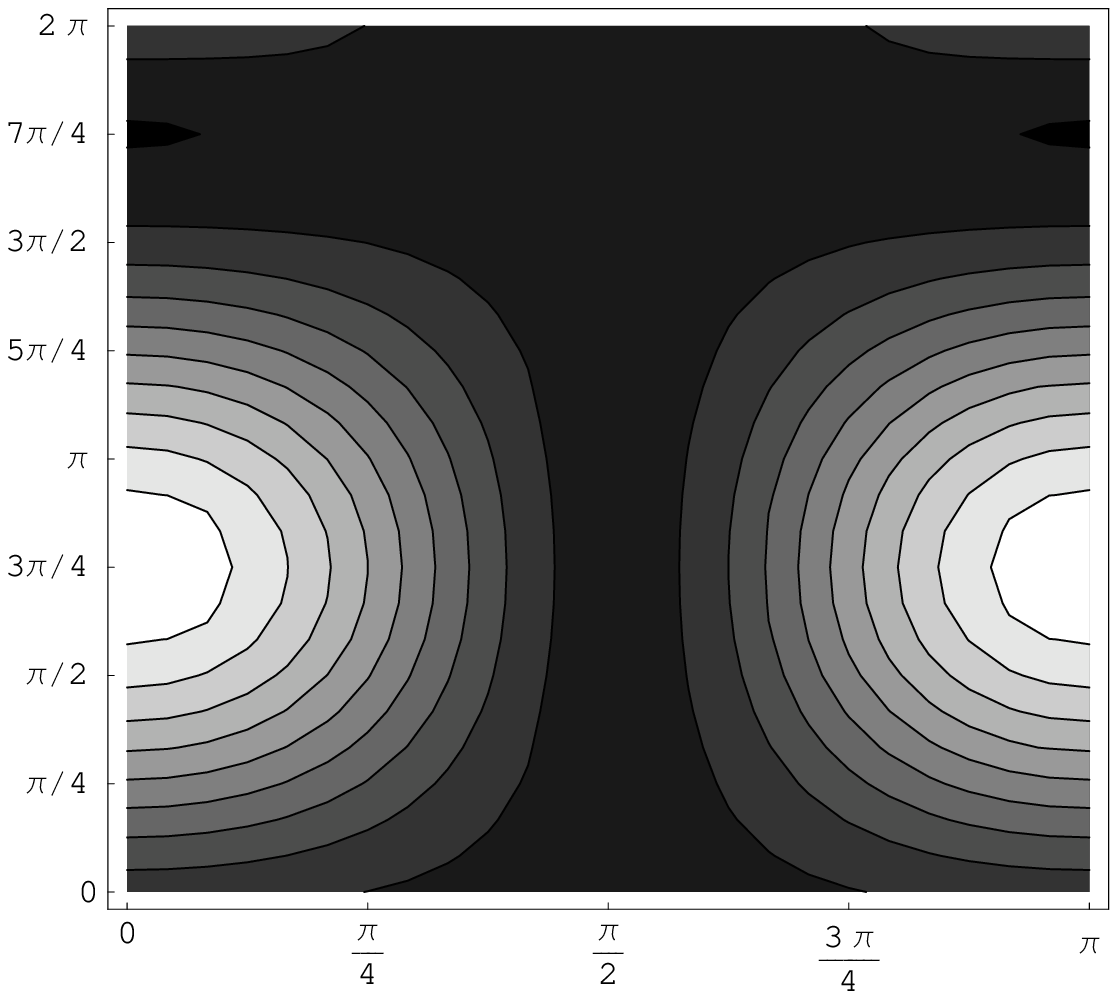}
        \put(-75,-7){\Large$\theta$}
             \put(-180,85){\Large$\phi$}
     \put(-180,160){$(b)$}~~\quad\quad\\
       \includegraphics[width=15pc,height=15pc]{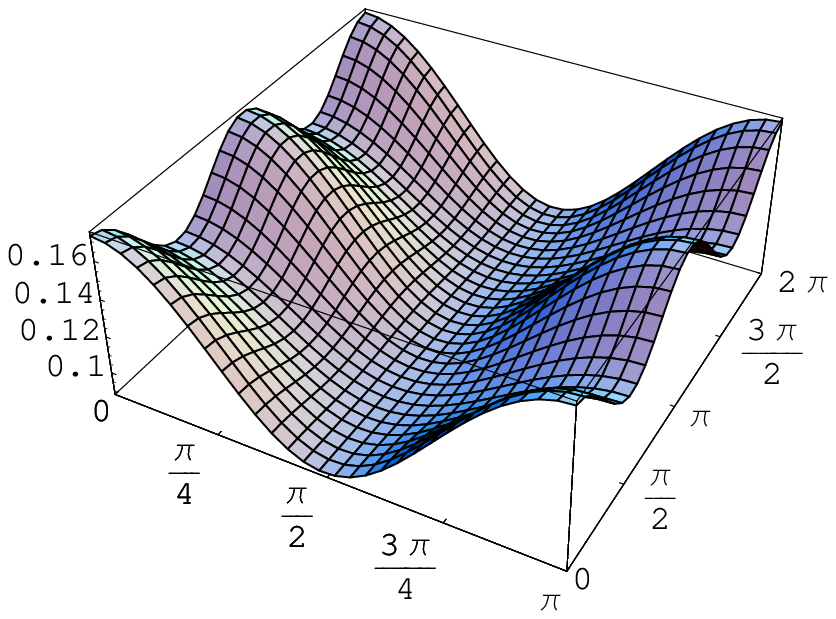}
       \put(-170,160){$(c)$}
       \put(-145,20){\Large$\theta$}
                \put(-15,55){\Large$\phi$}
               \put(-198,85){\Large$\mathcal{F}_\theta$}~~\quad\quad\quad
         \includegraphics[width=14pc,height=14pc]{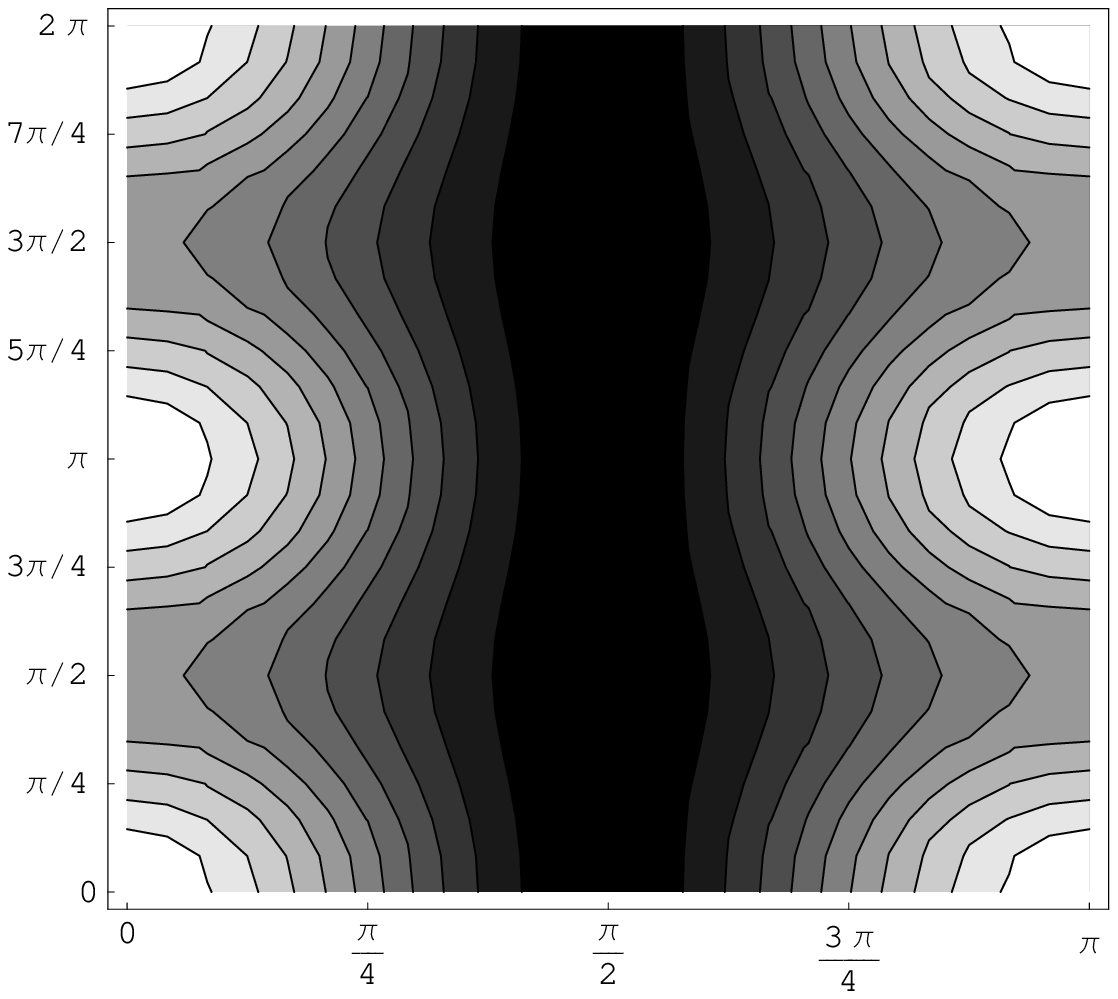}
              \put(-75,-7){\Large$\theta$}
             \put(-180,85){\Large$\phi$}
               \put(-175,160){$(d)$}
                    \caption{ (a)The Fisher information, $\mathcal{F}_\theta(\theta,\phi)$ at fixed
                    $r=\pi/8$, BSMA   for  system is prepared initially in
                    $\rho_{\psi^-}$,
                    (b)  WSMA for  system is prepared initially in $X$-state
                    with $c_{11}=-0.9,c_{22}=-0.8,c_{33}=-0.7$.}
\end{figure}

  Figs.($3c\&3d$) display the behavior of
$\mathcal{F}_\theta(\theta,\phi)$ WSMA for a system is initially
prepared in the $X$-state, where we set $c_{11}=-0.9,~
c_{22}=-0.8,~ c_{33}=-0.7$. In this case, the effect of phase
parameter, $\phi$ is different from that depicted in
Fig.$(3a\&3b)$, where $\mathcal{F}_\theta(\theta,\phi)$ reaches
its maximum values at $\phi=0,\pi,2\pi$. Moreover,
$\mathcal{F}_\theta(\theta,\phi)$ doesn't vanish  completely for
any value of $\phi\in[0,2\pi]$. The vanishing phenomena of the
Fisher information is due to the weight parameter, where it
decayes suddenly as $\theta$ increases. From Fig.$(3d)$, it is
clear that $\mathcal{F}_\theta(\theta,\phi)$ vanishes completely
at $\theta\in[3\pi/8,5\pi/8]$. For further values of $\theta$,
Fisher information re-birthes again to reach its maximum bounds at
$\theta=\pi$.

 From these Figs.(1-3), one {\it concludes}
that, the entanglement of the initial communication channel
between the users play  an important role on the teleported amount
of Fisher information, where the upper bounds of the teleported
Fisher information is large if the users use initially a maximum
entangled communication channel. The phenomena of the sudden decay
of Fisher information  is depicted WSMA as Unruh acceleration
increases. Moreover, the sudden increasing pheneomena of the
Fisher information is displayed for large values of Unruh
acceleration BSMA,  while the gradually increasing behavior is
displayed BSMA.

Moreover, the degree of estimating the teleported weight parameter
$\theta$, depends on the type of information which is encoded on
the teleported state. It is clear that, at $\theta=0$ or $\pi$,
$\mathcal{F}_\theta(\theta,\phi)$ is maximum, where at these
values  the initial  teleported state reduces to be
$\ket{\psi}=\ket{0}$ and $\ket{\psi}=e^{i\phi}\ket{1}$,
respectively, which means that the state carries only classical
information. On the other hand,  the minimum values of the
estimation degree appears at $\theta=\pi/2$, where the initial
teleported state is defined by
$\ket{\psi}=\frac{1}{\sqrt{2}}(\ket{0}\mp e^{i\phi}\ket{1})$,
which means that the initial state carries quantum information.

\begin{figure}[t!]
\centering
                       \includegraphics[width=15pc,height=15pc]{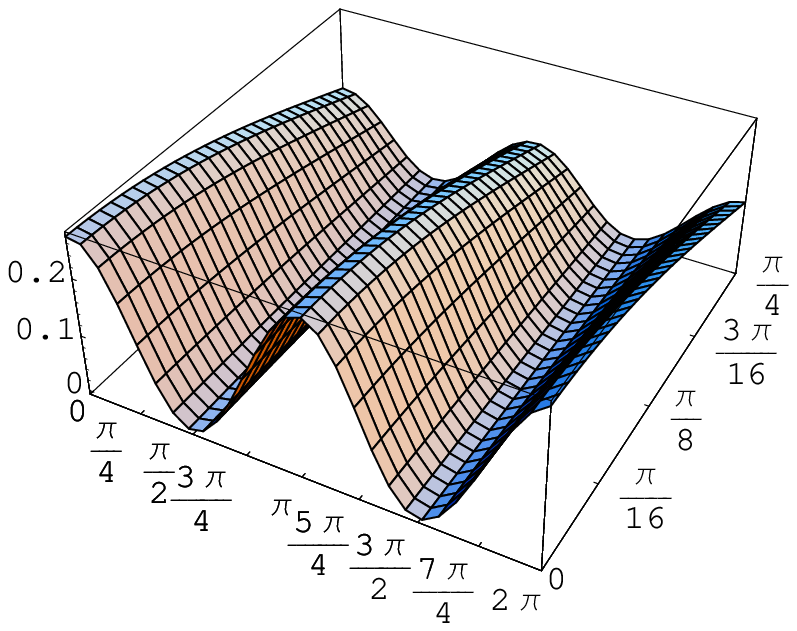}
           \put(-140,20){\Large$\phi$}
             \put(-16,60){\Large$r$}
      \put(-198,85){\Large$\mathcal{F}_\phi$}
     \put(-170,160){$(a)$}~~\quad\quad
          \includegraphics[width=15pc,height=15pc]{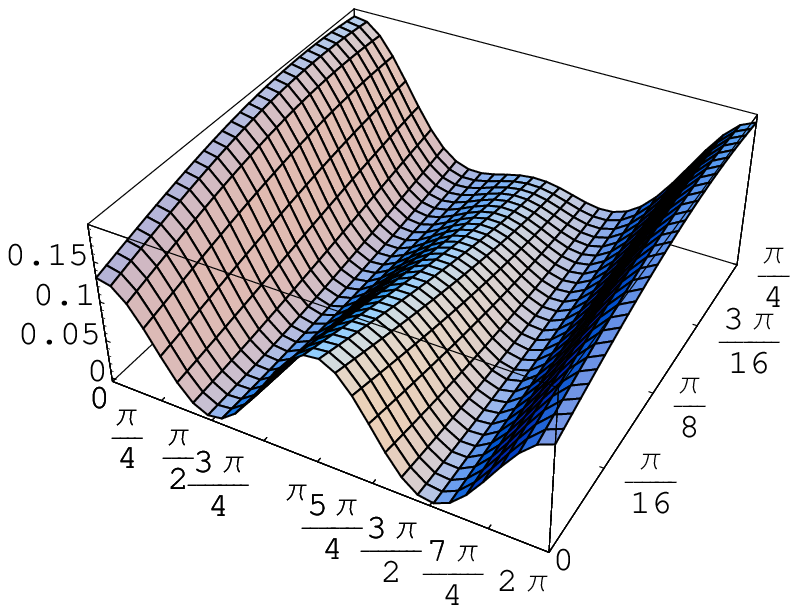}
          \put(-140,20){\Large$\phi$}
             \put(-16,60){\Large$r$}
     \put(-198,85){\Large$\mathcal{F}_\phi$}
      \put(-160,160){$(b)$}\\
     \includegraphics[width=15pc,height=15pc]{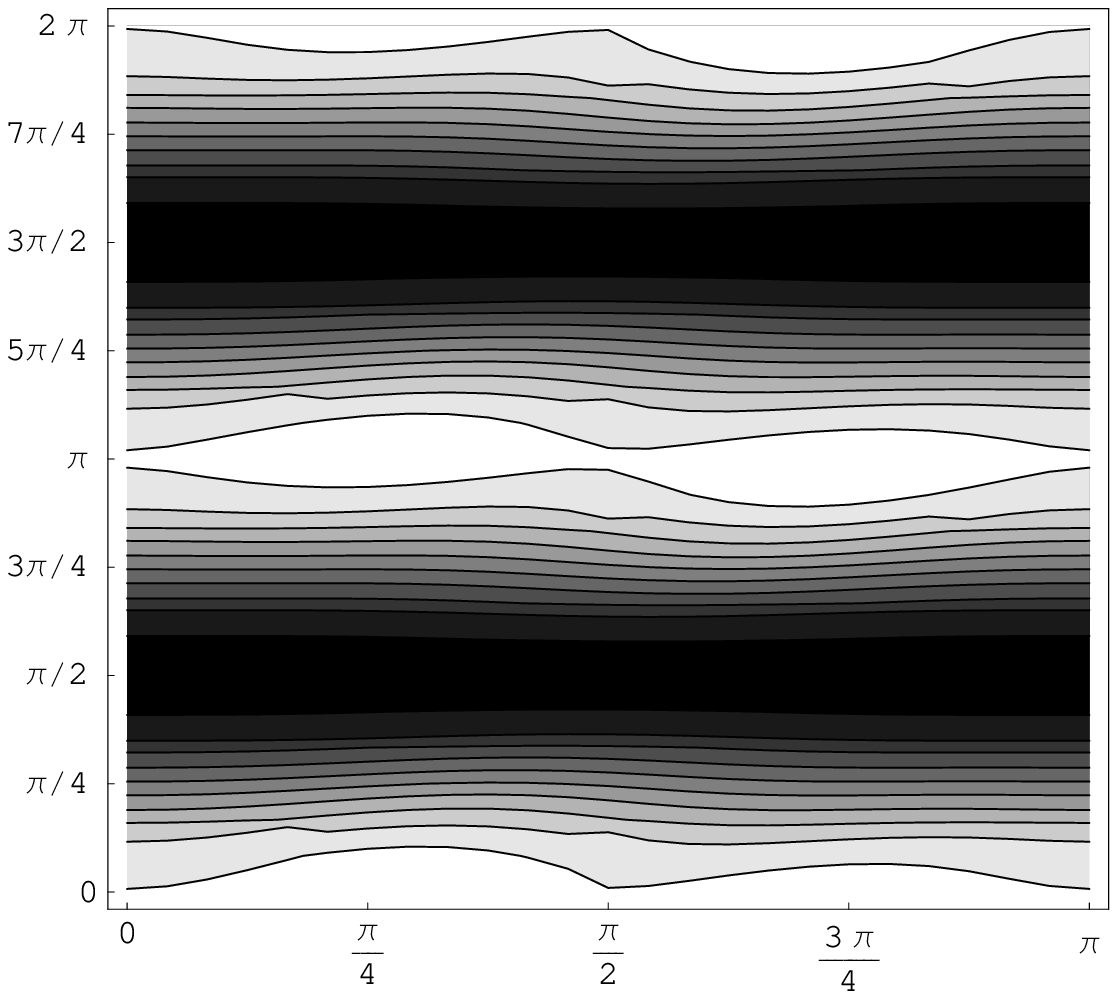}
             \put(-75,-7){\Large$\theta$}
                 \put(-185,90){\Large $\phi$}
     \put(-185,160){$(c)$}~~\quad\quad
             \includegraphics[width=15pc,height=15pc]{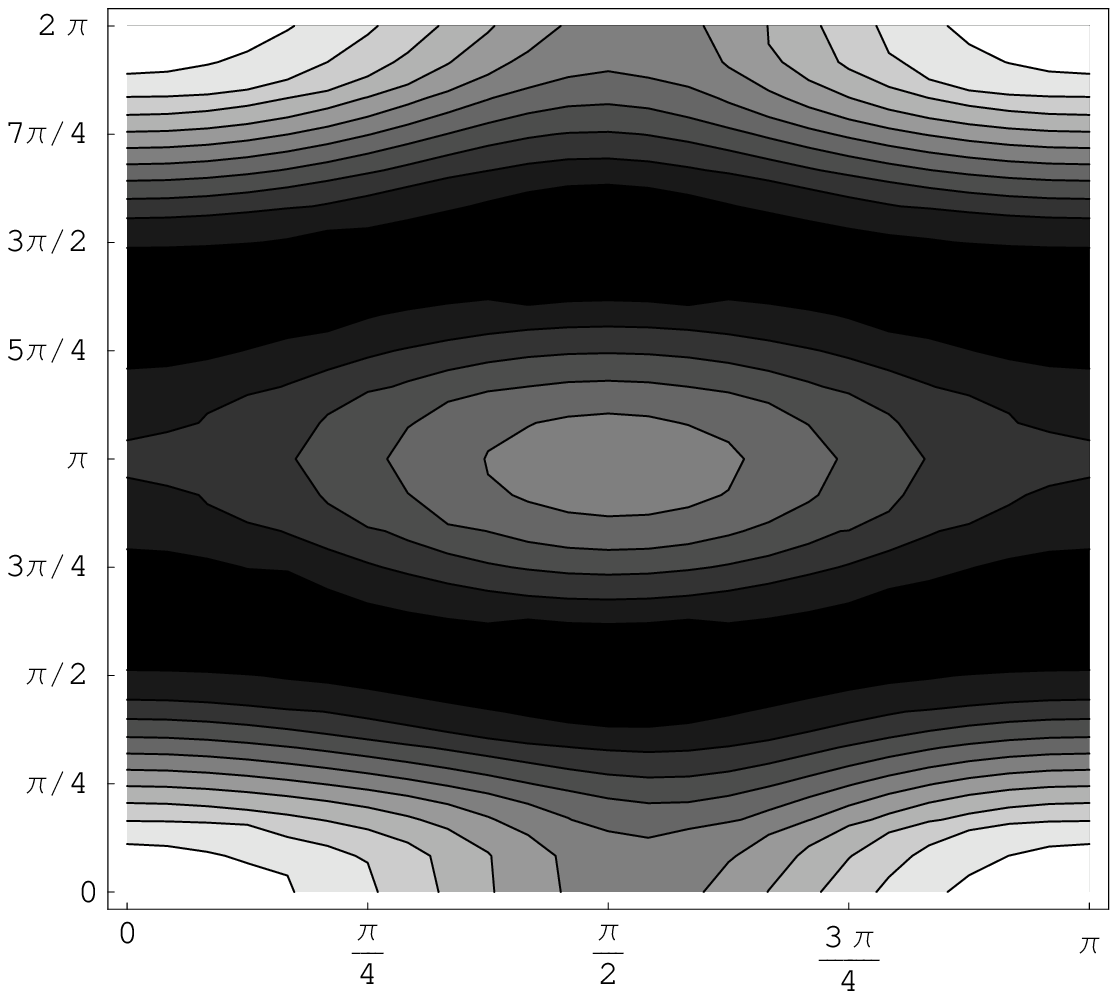}
           \put(-75,-7){\Large$\theta$}
              \put(-185,85){\Large$\phi$}
     % \put(-198,85){\Large$\mathcal{F}_\phi$}
      \put(-185,160){$(d)$}
            \caption{ The teleported  Fisher information $\mathcal{F}_{\phi}\phi,r)$, at fixed $\theta=\pi/4$
             with respect to the parameter
            $\phi $, for (a) WSMA i.e.,$q_L=1$, and (b)
            BSMA, with    $q_L=q_r=\frac{1}{\sqrt{2}}$. The Fisher information
            $\mathcal{F}_\phi(\phi,\theta)$at fixed $r=\pi/8$ for
            (c)WSMA and (b) BSMA.}
\end{figure}

\subsection{Estimating the phase  parameter, $\large\phi$}
The behavior of Fisher information, $\mathcal{F}_\phi(\phi,r)$ at
a fixed value $\theta=\pi/4$ with respect to the phase parameter
is shown in Fig.(4). Figs.(4a) and (4b) display the effect of the
Unruh acceleration on the teleported Fisher information
within/beyond the single mode approximation, respectively. The
general behavior shows that, $\mathcal{F}_\phi(\phi,r)$ decreases
as $r$ increases. The phenomena of the sudden changes of Fisher
information appears within/beyond the single mode approximation.
Within the single mode approximation, the upper bounds of the
Fisher information $\mathcal{F}_\phi(\phi,r)$ is larger than that
depicted for the beyond single mode approximation.

\begin{figure}[t!]
\centering
           \includegraphics[width=16pc,height=16pc]{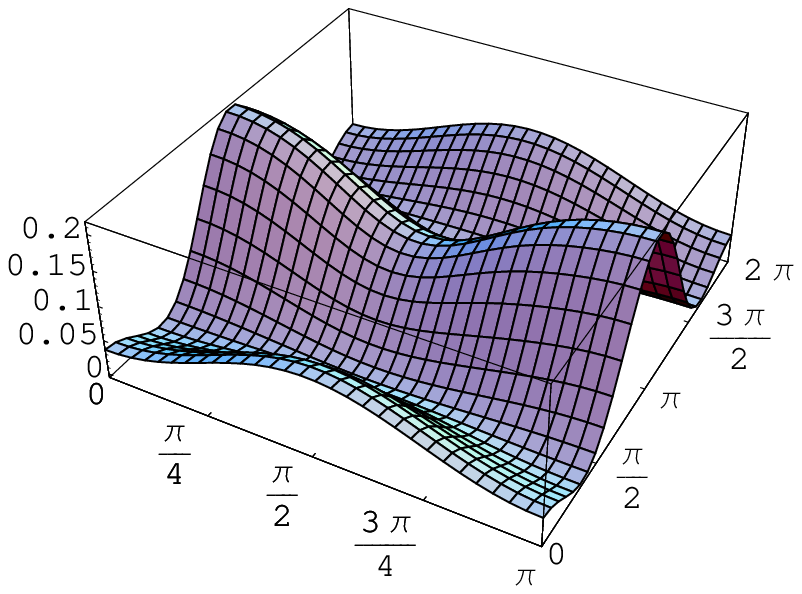}
           \put(-140,20){\Large$\theta$}
             \put(-16,60){\Large$\phi$}
      \put(-210,85){\Large$\mathcal{F}_\phi$}
     \put(-170,160){$(a)$}~~\quad\quad\quad
                   \includegraphics[width=15pc,height=15pc]{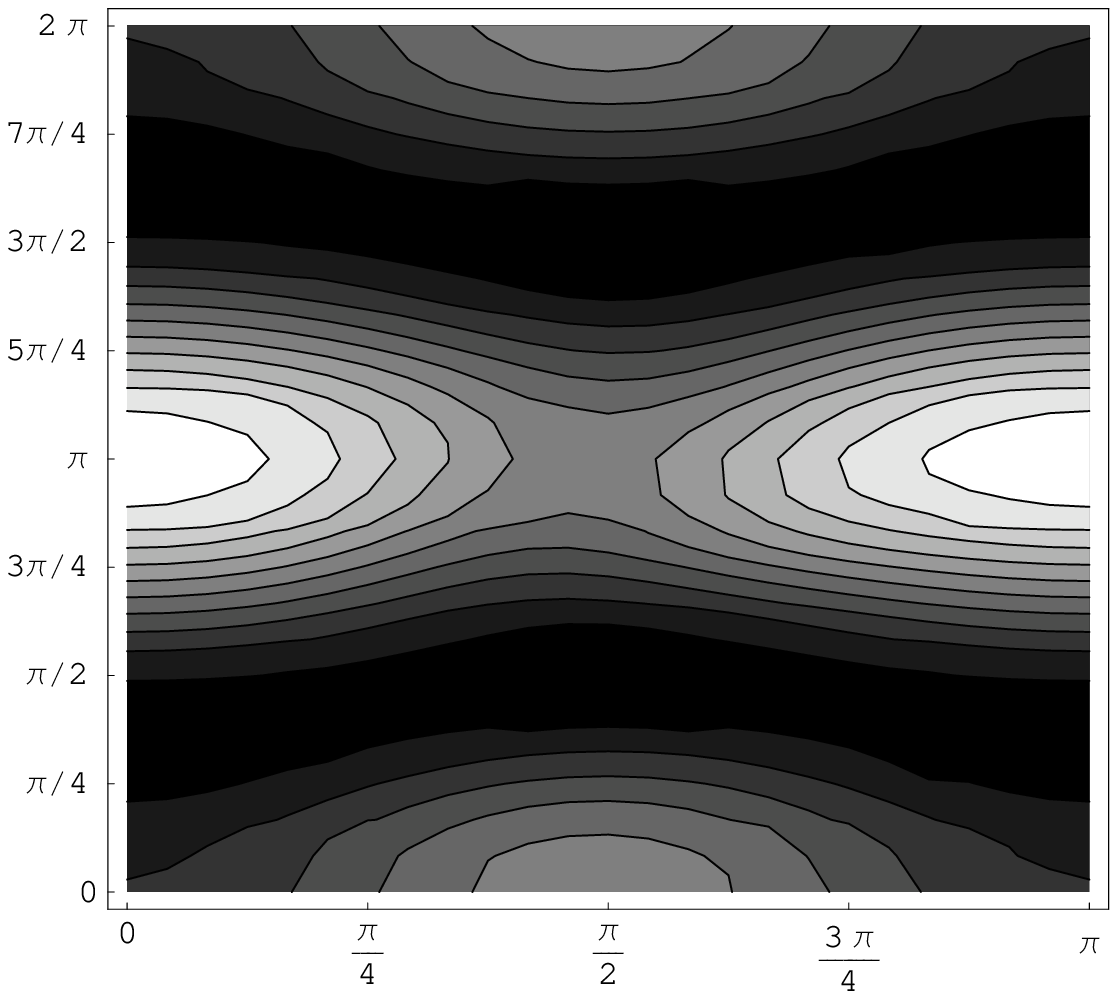}
                  \put(-75,-7){\Large$\theta$}
             \put(-180,90){\Large$\phi$}
     \put(-190,165){$(b)$}\\
         \includegraphics[width=16pc,height=16pc]{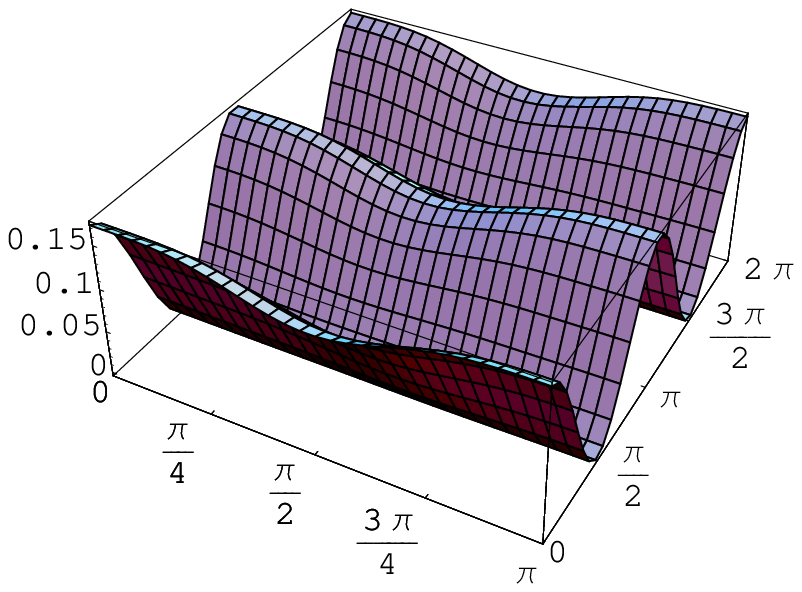}
 \put(-140,20){\Large$\theta$}
             \put(-16,60){\Large$\phi$}
      \put(-210,85){\Large$\mathcal{F}_\phi$}
     \put(-170,160){$(c)$}~~\quad\quad\quad
       \includegraphics[width=15pc,height=15pc]{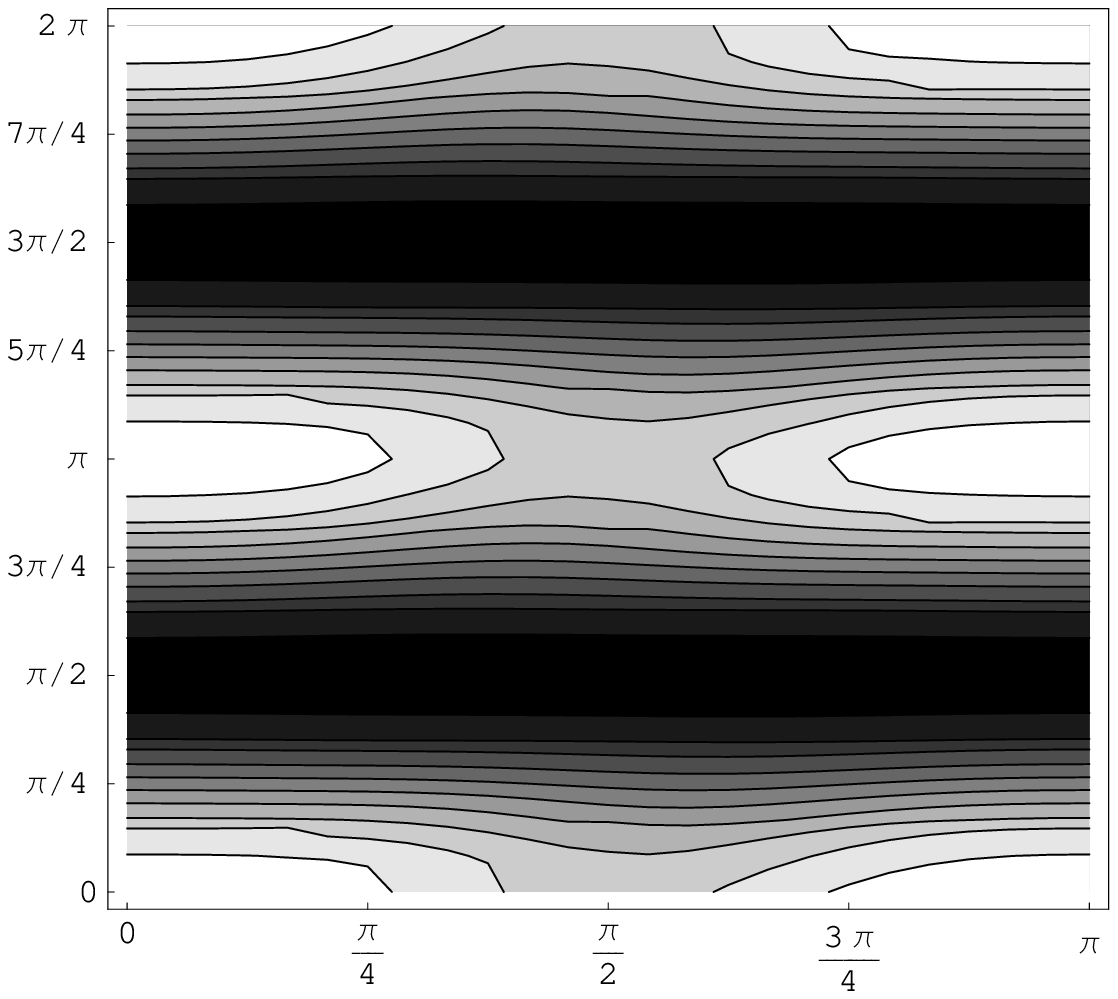}
            \put(-75,-7){\Large$\theta$}
             \put(-180,90){\Large$\phi$}
              \put(-190,165){$(d)$}
      \caption{ The dynamics of the Fisher information, $\mathcal{F}_\phi(\theta,\phi)$ at fixed $r=\pi/8$,
      (a) BSMA
      for  system is prepared initially in $\rho_{\psi^-}$, and
                    ( b)  WSMA for  system is prepared initially in the $X$-state
                    with $c_1=-0.9,c_2=-0.8,c_3=-0.7$.}
\end{figure}

Figs.$(4c\&4d)$ show the behavior of
 $\mathcal{F}_\phi(\phi,\theta)$ at a fixed value of the Unruh acceleration, $r=\pi/8$, within/beyond
the single mode approximation, respectively. In Fig (4c), the
weight parameter's effect decreases as $\phi$ increases. The
maximum values of  $\mathcal{F}_\phi(\phi,\theta)$ are depicted at
$\phi=\pi$ and $2\pi$, while the weight parameter can be
arbitrary.

  This effect is dramatically changed in the presence of the BSMA. It is evident that, at small values
of $\phi$,  $\mathcal{F}_\phi(\phi,\theta)$, decreases  as $\theta
$ increases to reach its minimum  value for the first time at
$\theta=\pi/2$. For further values of $\theta$,
$\mathcal{F}_\phi(\phi,\theta)$  increases gradually to reach its
maximum values at $\theta=\pi$. The effect of the weight parameter
is completely disappear as $\phi$ increases. However, as $\phi$
increases in the interval $[\pi/2,3\pi/4]$, and arbitrary values
of $\theta$, the Fisher information
$\mathcal{F}_\phi(\theta,\phi)$ almost vanishes completely.
However, for further values of $\phi$,
$\mathcal{F}_\phi(\theta,\phi)$ increases gradually to reach its
maximum values at $\phi=\pi$. This behavior is repeated again at
larger values of $\phi$.

%From Fig.(4), one can conclude that, the phenomena of the sudden
%decay and sudden changes of the Fisher information is due to  the
%leverage effect of the  of Unruh acceleration. The maximum values
%are predicted at $\phi=0,\pi,2\pi $.

Figs.($5a\&5b$) are devoted to investigate the behavior of
$\mathcal{F}_\phi(\theta,\phi)$ at a fixed value of $r=\pi/8$ for
a system is initially prepared in the singlet state,
$\rho_{\psi^-}$. In Fig.(5a),  the general behavior shows that,
the Fisher information, $\mathcal{F}_\phi(\theta,\phi)$  increases
gradually as $\theta$ increases to reach its maximum values at
$\theta=\pi/2$. At the same time as $\phi$ increases the effect of
the weight parameter decreases gradually to vanish completely at
$\phi\simeq\pi/4$. The upper bounds of
$\mathcal{F}_\phi(\theta,\phi)$ are depicted at $\phi=\pi$. These
maximum values decrease as $\theta$ increases to reach its minimum
value at $\theta=\pi/2$.  Fig(5b) shows clearly  the values of
$\theta$ and $\phi$ which maximize and minimize the Fisher
information.

The dynamics of the Fisher information of a teleported state by
using a communication  accelerated channel (within a single mode
approximation), initially prepared in the  $X$-state is displayed
in Figs.$(5c\&5d)$. It is clear that, as $\theta$ increases,
$\mathcal{F}_\phi(\theta,\phi)$ decays to reach its minimum value
 at $\theta=\pi/2$. The effect of the weight parameter decreases
as $\phi$ increases. On the other hand, the sudden changes of the
Fisher information,$\mathcal{F}_\phi(\theta,\phi)$ are depicted as
$\phi$ increases.

\subsection{Estimation of Unruh acceleration}
\begin{figure}[t!]
\centering
                       \includegraphics[width=15pc,height=15pc]{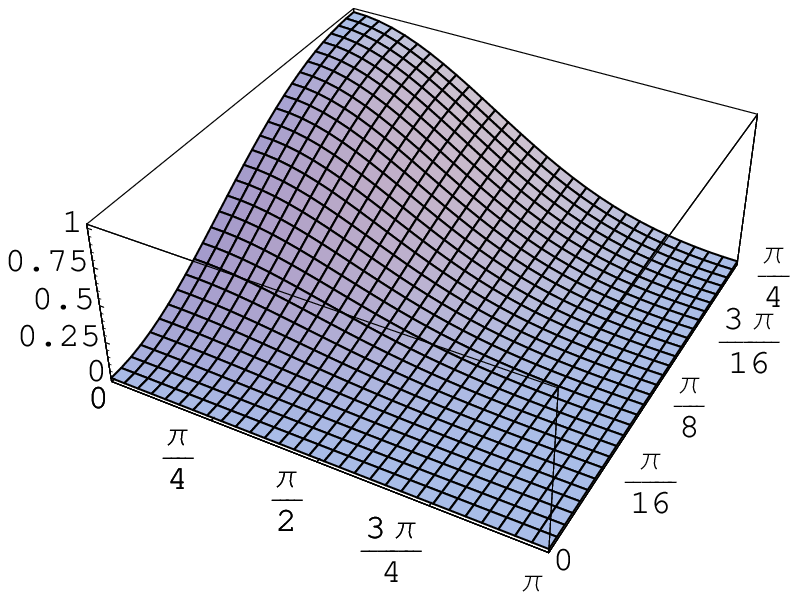}
           \put(-140,20){\Large$\theta$}
             \put(-16,60){\Large$r$}
      \put(-198,85){\Large$\mathcal{F}_r$}
     \put(-170,160){$(a)$}~~\quad\quad
            \includegraphics[width=15pc,height=15pc]{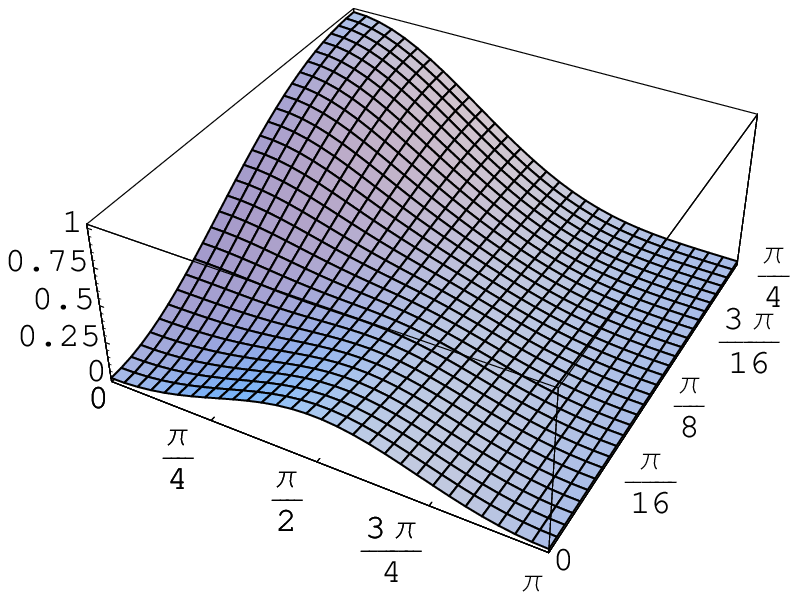}
          \put(-140,20){\Large$\theta$}
             \put(-16,60){\Large$r$}
     \put(-198,85){\Large$\mathcal{F}_r$}
      \put(-160,160){$(b)$}\\
     \includegraphics[width=15pc,height=15pc]{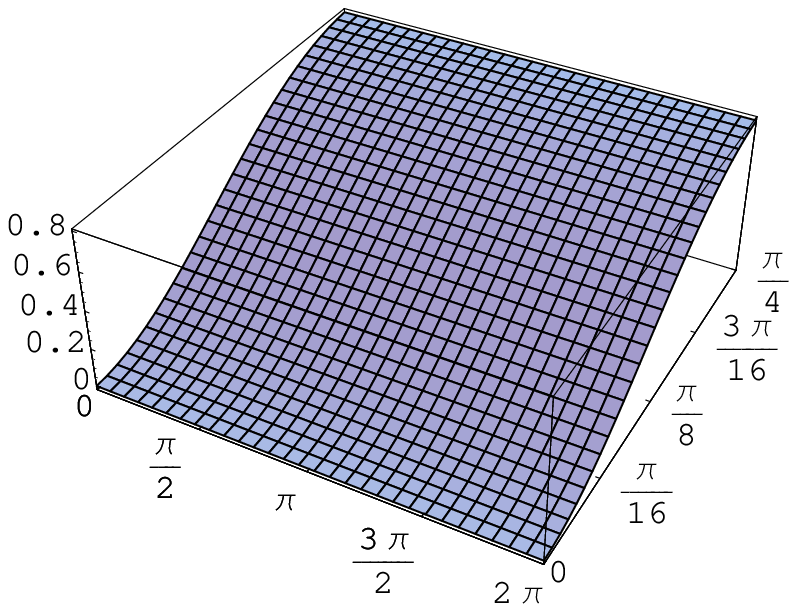}
     \put(-140,20){\Large$\phi$}
             \put(-16,60){\Large $r$}
      \put(-198,85){\Large$\mathcal{F}_r$}
     \put(-170,160){$(c)$}~~\quad\quad
            \includegraphics[width=15pc,height=15pc]{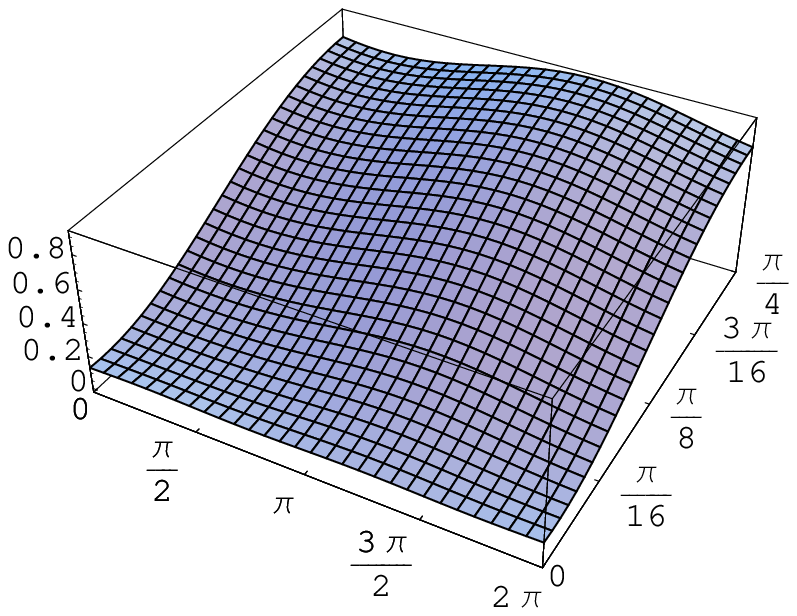}
                 \put(-140,20){\Large$\phi$}
             \put(-16,60){\Large$r$}
      \put(-198,85){\Large$\mathcal{F}_r$}
      \put(-160,160){$(d)$}
            \caption{ The  Fisher information $\mathcal{F}_{r}(\theta,r)$, at fixed value of $\phi=\pi/4$ for a  system is initially prepared in
          the Bell state,  $\rho_{\phi^+}$, for The figures  (a) WSMA (b) BSMA.
          Figs(c,d) represent the Fisher information, $\mathcal{F}_{r}(\phi,r)$ at a fixed value of $\theta=\pi/4$
          where (c)WSMA and, (b) BSMA. }
\end{figure}

It is clear that,  the final teleported state(8) not only  depends
on  the initial parameter but also on the Unruh acceleration
parameter. Therefore, it is important to estimate this parameter
by quantifying the amount of Fisher information with respect to
the Unruh acceleration, $\mathcal{F}_r(\theta,r)$. In Fig(6), we
investigate the effect of the initial  parameters $\theta $ and
$\phi$ on the behavior $\mathcal{F}_r(\theta,r)$
 within/beyond the single mode approximation. The effect of the
 parameter $\theta$ is displayed in Figs.(6a$\&6b)$, where we set
 $\phi=\pi/4$. It is manifest that, $\mathcal{F}_r(\theta,r)$ increases as
 $r$ increases to reach its upper bounds at $\theta=0$. For
 further values of $\theta$, the Fisher information $\mathcal{F}_r$ decreases to vanish
 completely. For any arbitrary value of $\theta$ and small values
 of $r$, $\mathcal{F}_r$ is almost zero. However, this behavior is
 changed if  the beyond single mode approximation  is considered (Fig.(6b)), where
 $\mathcal{F}_r$ increases gradually  as $\theta$ increases to reach its maximum values
 $\theta=\pi/2$. For further values of $\theta$, Fisher
 information $\mathcal{F}_r$ decreases gradually to disappear at
 $\theta=\pi$.

The effect of the phase  parameter  on $\mathcal{F}_r(\phi,r)$ is
described in Fig.($6c\&6d$), where we set $\theta=\pi/4$. It is
clear that, within the single mode approximation,
$\mathcal{F}_r(\phi,r)$ increases as $r$ increases, where the
phase parameter has a feeble effect. On the other hand, beyond the
single mode approximation $\mathcal{F}_r(\phi,r)$ increase
gradually as $\phi$ increases and  the maximum values are reached
at $\phi=3\pi/2$. The upper bounds of Fisher information that
depicted in Fig.$(5c\&5c$) is smaller than that shown in
Fig.(5a$\&5b$).

%From Fig.(6), one can conclude that, the gained parameter $r$, can
%be estimated with high precision if the teleported state encodes
%only classical information.

\section{Conclusion}
In this contribution, we investigate the possibility of estimating
the initial teleported parameters and the gained parameters during
the teleportation process. The partner, Alice and Bob, share
initially a communication channel  of self-transposed class, which
could be maximum  entangled Bell state , or $X$-state. It is
considered that, only Bob'squbit is accelerated while Alice's
qubit is in the inertial frame. The final accelerated state
between Alice and Bob is used to teleport unknown state from Alice
to Bob by means of the teleportation protocol. The final
teleported state  depends on the initial parameters in addition to
the Unruh acceleration parameter. Fisher information is used a
measure of estimating the initial and the gained parameters, where
we calculate it corresponding to each parameter within and beyond
single mode approximations.

The possibility of estimating  the teleported parameters
within(beyond) the single mode approximation decreases/increases
as the Unruh acceleration increases. The maximum values of
estimation depend on the estimated parameter and the approximation
mode. For estimating the teleported weight  parameter at a
particular value of the Unruh acceleration, the phase parameter
has a slightly effect within the single mode approximation, while
this effect is large beyond the  single mode approximation. A
similarly behavior is depicted when the teleported phase parameter
is teleported, namely the weight parameter has a slight/large
effect within/beyond single mode approximation, respectively.

Estimating the gained parameter,(Unruh acceleration) is discussed
within/beyond the single mode approximation. It is clear that, the
degree of estimation increases as the weight parameter increases
to reach its maximum value, when the initial teleported state
encode only classical information. For small values of
acceleration, the weight parameter, the phase  parameter have  a
slightly effect on the degree of estimating the acceleration
parameter within the single mode approximation. The maximum value
of estimating the gained parameter for arbitrary weight parameter
is larger than that depicted for the arbitrary value of the phase
parameter.

The effect of different cases of the initial states also, is
discussed, where we show that using different classes of Bell
states cause a shift of the maximum and minimum bounds of the
Fisher information.  Moreover, there are some extra tops  appear
for the singlet state. The degree of estimating  of the teleported
parameter by using $X-$ state is similar to that predicted for the
singlet state but with smaller upper bounds, where the degree of
estimation depends on the initial entanglement of the
communication channel.

In {\it conclusion}, it is possible to estimate the teleported and
the gained parameters by means of Fisher information. The maximum
values of the  estimation degree depend on the used approximation,
the entanglement of the initial state between the partners and the
structure of the initial teleported state. One can estimate these
 parameters with a large probability if the users teleported a
classical information.

\end{document}